# Manipulating Semicrystalline Polymers in Confinement

*Nitin Shingne[1], Markus Geuss,[2,5], Thomas Thurn-Albrecht[1], Hans-Werner Schmidt[3], Carmen Mijangos[4], Martin Steinhart[*,5], Jaime Martín[*,6,7]*

[1] Institute of Physics, Martin Luther University Halle-Wittenberg, Heinrich-Damerow-Str. 4, D-6120 Halle, Germany

[2] L'institut iPrint, Haute École d'ingénierie et d'architecture Fribourg, Route de l'Ancienne Papeterie 180 ▪ Case postale 146, Marie Sklodowska Marly 1, Switzerland

[3] Macromolecular Chemistry I, Bavarian Polymer Institute, and Bayreuth Center for Colloids and Interfaces, University of Bayreuth, 95440 Bayreuth, Germany

[4] Instituto de Ciencia y Tecnología de Polímeros, CSIC. Juan de la Cierva 3, Madrid 28006, Spain

[5] Institut for Chemistry, University of Osnabrück, Barbarastr. 7, 49076 Osnabrück, Germany

[6] POLYMAT, University of the Basque Country UPV/EHU Avenida de Tolosa 72, 20018 Donostia-San Sebastián, Spain.

[7] Centre for Plastic Electronics and Department of Materials, Imperial College London, Exhibition Road, London, SW7 2AZ, UK

**Autor Information**

Corresponding authors:

*Email: jaime.martin@polymat.eu
*Email: martin.steinhart@uos.de

**Abstract**

Because final properties of nanoscale polymeric structures are largely determined by the solid-state microstructure of the confined polymer, it is imperative not only to understand how the microstructure of polymers develops under nanoscale confinement but also to establish means to manipulate it. Here we present a series of processing strategies, adapted from methods used in bulk polymer processing, that allow to control the solidification of polymer nanostructures. Firstly, we show that supramolecular nucleating agents can be readily used to modify the crystallization kinetics of confined poly(vinylidene fluoride) (PVDF). In addition, we demonstrate that microstructural features that are not traditionally affected by nucleating agents, such as the orientation of crystals, can be tuned with the crystallization temperature applied. Interestingly, we also show that high crystallization temperatures and long annealing periods induce the formation of the γ modification of PVDF, hence enabling the simple production of ferro/piezoelectric nanostructures. We anticipate that the approaches presented here can open up a plethora of new possibilities for the processing of polymer-based nanostructures with tailored properties and functionalities.

## 1. Introduction

With the rapid development of nanomedicine as well as organic optoelectronics and photonics, the demand for well-defined sub-micron scale polymeric structures – i.e. polymer nanostructures – is rapidly rising. Polymer molecules in these material systems are confined to volumes of the same order to magnitude as the characteristic length scales of most of the structure-development processes occurring in polymers, the most important of which is the crystallization. This spatial restriction may cause the crystallization process to develop differently than under bulk conditions. Polymer nanostructures often exhibit different structural properties than unconfined bulk counterparts [1-4] and, consequently, they also have different functional properties, for example, a lower thermal conductivity[5] or modified electrical behavior[6]. It is thus imperative not only to understand how nanoscopic morphologies of polymers develops under confinement but also to establish means to manipulate it.

The addition of additives aiding the nucleation of polymer crystals, i.e. the so-called nucleating agents, is the primary route to control bulk crystallization – e.g. crystal size, density and selection of polymorphs – in industrial processing. These additives act as heterogeneous nuclei with epitaxial surfaces reducing the free energy of nucleation, which results in a smaller critical nucleus size required for polymer crystallization, and eventually in the development of smaller crystals [7-8]. Recently, the approach has been further adapted to more advance processing methods, such as those of organic semiconductors[9]. Thus, inspired by this versatility, we aimed to explore whether this simple, yet powerful strategy can be applied to manipulate the crystallization and the structural properties of polymeric nanostructures. To this end, we selected the N,N'N''-tris-1,2-dimethylpropyl-1,3,5-bezenetricarboxamide (BTA, Figure 1a), a supramolecular nucleating agent recently introduced for isotactic polypropylene (i-PP)[8], which has also demonstrated activity for poly(vinylidene fluoride) (PVDF)[10] (Figure 1a). When mixed with bulk polymers at temperatures high enough, BTA species dissolve in the molten polymer. Then, upon cooling BTA is prone to one-dimensional self-assembly and crystallization in needle-like structures on which the polymer starts to crystallize.

Herein we show that supramolecular nucleating agents can be readily used to modify the crystallization of polymer nanostructures through the variation of the crystallization kinetics. Furthermore, we demonstrate that the morphological features that are commonly not sensitive to nucleating agents, such as the orientation of crystals, can be controlled by the temperature at which isothermal crystallization is carried out. Interestingly, we also show that high crystallization temperatures and long annealing periods induce the formation of the γ modification of PVDF, hence enabling the production of ferroelectric/piezoelectric one-dimensional nanostructures.

## 2. Methods

For the study of the nucleating activity on PVDF nanotubes, the solid mixture of PVDF (110 kDa with PDI of 2.8) and BTA was placed on the surface of anodic aluminium oxide (AAO) nanoporous templates with average pore diameter of 400 nm and a pore depth of 100 μm[11] at a temperature above the dissolution of the BTA in the molten polymer (T=265 ºC). Under these conditions, the molten polymer wets the AAO nanopore walls in the complete regime[12], which results in the formation of liquid polymeric precursor films on the pore walls. Since the pore radius is larger than thickness of the precursor film, tubular precursor films are obtained – i.e. molten polymer nanotubes, which can be thereafter solidified to achieved PVDF nanotubes. This solidification was accomplished by quenching in ice-water in order to avoid migration of nucleating agents out of the pores during cooling. Finally, the residual PVDF films located on top of the AAO hard templates were removed with scalpel so that the nanotubes within the AAO were isolated entities separated from each other. To erase the thermal history of the infiltrated AAO templates, they were again molten before the crystallization experiments were carried out. Reference neat PVDF nanotube samples were treated in the same way. The crystallization temperatures and kinetics of polymer nanostructures were investigated using a TA Instrument (New Castle, DE) Q100 differential scanning calorimeter (DSC) under $N_2$ atmosphere.

Isothermally crystallized samples for structural analysis were prepared employing the procedure above indicated, followed by an isothermal step which allowed the samples to be crystallized for the time *t*. The PVDF used for this study had a $M_w$ of 180 kDa and a PDI of 2.4. The structural characterization of these samples was accomplished by wide-angle X-ray scattering using a PANalytical X'pert diffractometer. Further details on the nanotube preparation and the characterization methods are provided in the Supporting Information.

## 3.    Results and Discussion

Self-ordered anodic aluminum oxide (AAO)[11] (Figure 1b) containing arrays of aligned cylindrical nanopores with rigid walls is an ideal matrix system to elucidate whether nucleating agents are effective to alter the crystallization of confined polymers. AAO templates allow confining crystallizable molten polymers into well-defined nanoscale volumes. The thermal stability allows application of suitable isothermal and non-isothermal temperature profiles; moreover, the anisotropy of the AAO nanopores with high aspect ratios permits assessing which crystal faces are preferentially oriented normal to the AAO nanopore axes. Thus, PVDF nanotubes (Figure 1a) with an outer outer diameter of 400 nm, a wall thickness of approx. 35 nm and a length of 100 μm were fabricated by wetting AAO templates under conditions promoting the infiltration of molten polymeric precursor films within the nanopores and subsequent solidification[12-14]. Figures 1b and 1c show scanning electron microscopy (SEM) images of the AAO templates and PVDF thus nanotubes produced, respectively. The concentration of the BTA used for our study was optimized to 1200 ppm (see Supporting Information Table S1).

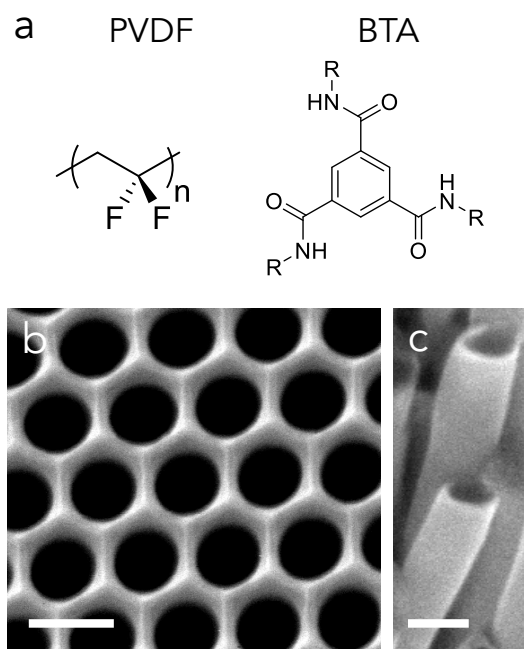

**Figure 1**. (a) Molecular structure of poly(vinylidene fluoride) (PVDF) and 1,3,5-bezenetricarboxamide (BTA), where −R represent 1,2-dimethylpropyl substituents. SEM images of (b) the surface of a self-ordered AAO template and (c) PVDF nanotubes prepared by molding PVDF against AAO. The scale bars correspond to 500 nm.

First evidence that the crystallization of PVDF nanotubes can be modified with BTA nucleating agents was obtained when the non-isothermal crystallization experiments from the melt were conducted by differential scanning calorimetry (DSC). As shown in Figure 2a, an increase of the onset crystallization temperature by 6 ºC was recorded for the PVDF nanotubes crystallized with BTA as compared to the neat nanotubes (from 133 ºC to 139 ºC, see Table S1), which suggest that BTA increases the nucleation rate. Interestingly, a similar increase of crystallization temperature was detected for bulk PVDF, evidencing a similar efficiency of the BTA both in bulk and in confined PVDF (Supporting Information Table S1).

In order to further assess the modification of the crystallization kinetics of confined PVDF induced by the BTA, we performed isothermal crystallization experiments. DSC was employed to monitor the advance of the crystallization, which was analyzed using the Avrami model[15]. In the Avrami model, the advance of the crystallinity with time is expressed as $V_c(t) = 1 - exp(-Kt^n)$, where $V_c(t)$ is the crystallinity in terms of the volume fraction of crystals that develop within a specific time $t$; $K$ is the rate constant of the crystallization process, and $n$ is the Avrami exponent, which relates to the time dependency of the overall crystallization process and can be further interpreted as $n = n_n + n_g$, where $n_n$ and $n_g$ are the contributions to $n$ from the nucleation and crystal growth process, respectively. Figure 2b and 2c depicts the variation of the relative crystallinity (expressed as the normalized enthalpy values, $\Delta H$, grey circles) with time for (b) neat PVDF nanotubes (at $T_c$ = 143 ºC, 145 ºC, 147 ºC, 149 ºC and 151 ºC) and (c) PVDF nanotubes crystallized with BTA (at $T_c$ = 148 ºC, 150 ºC, 152 ºC and 154 ºC). The details of the Avrami analysis of bulk and PVDF nanostructures are provided in the Supporting Information (Figures S2 and S3).

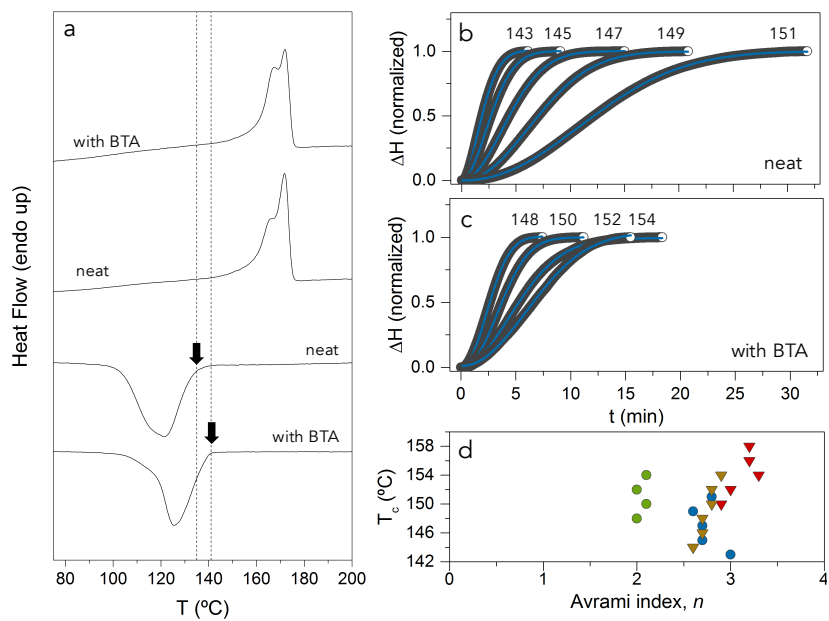

**Figure 2**. (a) Representative heating and cooling 2$^{nd}$ DSC scans for 400 nm in diameter PVDF nanotubes crystallized in the presence and in the absence of BTA (the latter denoted as "neat"). The dashed lines and the arrows indicate the onset crystallization temperatures. Variation of the relative crystallinity (expressed as the normalized enthalpy values, $\Delta H$, grey circles) with time for (b) neat PVDF nanotubes ($T_c$ = 143 ºC, 145 ºC, 147 ºC, 149 ºC and 151 ºC) and (c) PVDF nanotubes crystallized with BTA ($T_c$ = 148 ºC, 150 ºC, 152 ºC and 154 ºC). Blue solid lines represent fits to the Avrami equation. (d) Values of the Avrami index, $n$, plotted versus $T_c$ for nanotubes crystallized with BTA (green circles) and neat PVDF nanotubes (blue circles). $n$ values for bulk PVDF crystallized with BTA (red triangles) and neat bulk PVDF (orange triangles) are included for comparison (for detailed analysis of the bulk samples see the Supporting Information Figures S2 and S3).

Fits to the Avrami equation revealed different crystallization kinetics for PVDF nanotubes crystallized with and without BTA. For example, a different dependency of the crystallization process with time – expressed in terms of the exponent, $n$ – was deduced. Whereas for neat PVDF nanotubes $n$ amounted to ~3, $n$ values of ~2 were obtained for nanotubes with BTA (Figure 2d). This outcome can be rationalized by assuming a change of the nucleation mechanism due to addition of BTA as follows: The dimensionality of the crystal growth (which is reflected in $n_g$) can be expected to be identical in both samples because: *(i)* the growing crystals are subject to the same degree of two-dimensional geometric confinement imposed by the rigid walls of the AAO nanopores, and *(ii)* the nucleation centers can be assumed to be randomly distributed and oriented within the nanotubes. Hence, any differences in $n$ must result from a change in $n_n$, i.e. in nucleation kinetics. It is reasonable to assume that in the isolated neat PVDF nanotubes homogeneous nucleation events sporadically occur – with $n_n$~1, as has been previously suggested[16]. In BTA-containing PVDF nanotubes, however, PVDF crystal nuclei develop simultaneously on the surface of the pre-formed BTA particles so that nucleation does not contribute to the time dependency of the crystallization and thus $n_n \approx$ 0. This result means, moreover, that $n_g$ amounts to ~2 in the PVDF nanotubes irrespective of the presence or absence of BTA and that crystal growth would be two-dimensional. Moreover, a further evidence of the change of the nucleation mechanism in the PVDF nanotubes due to the addition of BTA is found in the notably lower apparent activation energy for the crystallization, $E_a$, of the BTA containing nanotubes ($E_a$ = 251 kJ/mol) compared to that of neat nanotubes ($E_a$ = 440 kJ/mol) (Figure S4).

Having established that crystallization kinetics of polymers in cylindrical confinement can be manipulated by nucleating agents, we wanted to further scrutinize our approach and evaluate whether the nucleating agents modified the degree of crystallinity and the orientation of the crystals in the PVDF nanotubes. The degree of crystallinity of neat PVDF nanotubes crystallized non-isothermally at a cooling rate of -1 ºC/min was found to be of ~38 % but decreased to ~32 % in the presence of BTA (Supporting Information Table S3). Both in the presence and the absence of BTA, the (020) lattice planes and, to a lesser degree, the (110) planes were oriented normal to PVDF nanotube axes in accordance with previous reports[6, 16] (Supporting Information Figure.S5). Clearly, the PVDF crystals in the PVDF nanotubes show pronounced orientation, but this feature is independent of the addition of BTA. Thus, our results evidence that, like in bulk polymers, nucleating agents do not significantly alter neither the degree of crystallinity nor the crystal texture of the PVDF nanotubes – at least when non-isothermal crystallizations are performed (Supporting Information Table S3 and Figure S5). The reason for this outcome is that crystal orientation and crystallinity are to a significant extent established during crystal growth. Thus, these features are largely influenced by the balance between the nucleation stage and the growth stage[17]. The isothermal crystallization of polymers at different temperatures has proven to be an efficient approach to manipulate that balance [17] and, hence, it seems a promising strategy to control the crystal orientation of confined polymers.

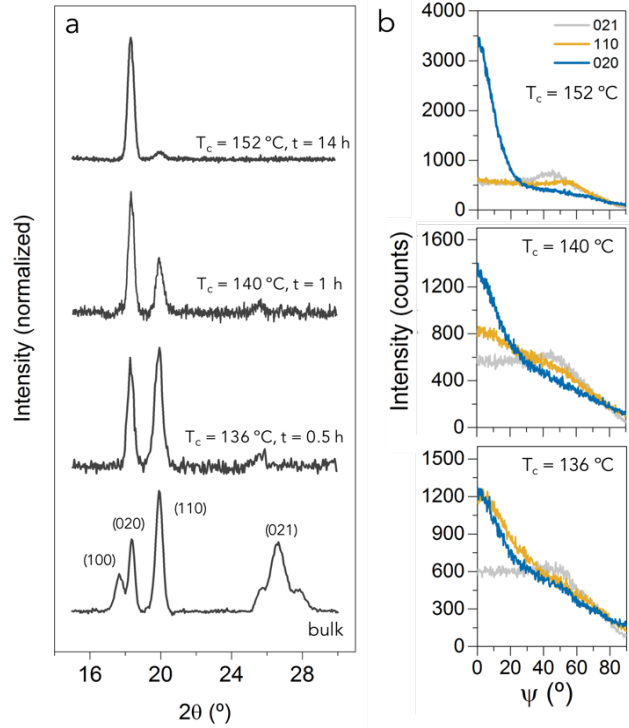

**Figure 3**. WAXS characterization of neat PVDF nanotubes located in aligned AAO nanopores. (a) WAXS θ/2θ patterns of neat bulk PVDF (at the bottom) and neat PVDF nanotubes isothermally crystallized at the indicated isothermal crystallization temperatures $T_c$. (b) Schulz scans of neat PVDF nanotubes crystallized at the indicated $T_c$: (020), blue; (110), orange; (021), grey.

Figure 3a shows the wide-angle x-ray scattering (WAXS) θ/2θ patterns for bulk neat PVDF and neat PVDF nanotubes isothermally crystallized at different $T_c$. The WAXS measurements were taken on PVDF nanotubes located in the aligned nanopores of self-ordered AAO. The reflections appearing in the WAXS patterns can be indexed as follows: (100) at 2θ = 17.9º; (020) at 2θ = 18.3º; (110) at 2θ = 20.0º; and (021) at 2θ = 26.6º [18], and correspond to α-PVDF ($a$ = 4.96 Å, $b$ = 9.64 Å and $c$ = 4.62 Å).[18-19] Given the geometry of the WAXS set-up that we used (Supporting Information Figure S6) detected scattering intensity exclusively originates from sets of lattice planes oriented perpendicular to the long axis of the AAO nanopores and the neat PVDF nanotubes, two main outcomes can be highlighted from these data:

*(i)* The diffraction peak at 26.6° present in bulk PVDF, corresponding to (021) lattice planes, is absent in all WAXS patterns of the PVDF nanotubes. This indicates that no (021) planes are oriented normal to the long axes of the AAO nanopores and the PVDF nanotubes. The absence of the (021) diffraction peak is in line with previous findings [16, 20] and can be explained by the fact that the lamellae with non-zero *l*-index cannot grow along the PVDF nanotubes as their directions of fast crystal growth is inclined with respect the PVDF nanotube axes[16].

*(ii)* The (020)/(110) intensity ratio increases progressively as $T_c$ increases, revealing that the selection of <*hkl*> directions with non-zero *l*-index aligned with long axis depends strongly on the $T_c$. Whereas at low $T_c$ a significant proportion of the PVDF crystals grows normal to the (110) plane, at high $T_c$ the PVDF crystals nearly exclusively grow normal to the (020) planes.

In order to further analyze the crystal orientations in the neat PVDF nanotubes, we measured Schulz scans for the (110), (020) and (021) reflections[21-22] and for $T_c$s of 136 ºC, 140 ºC and 152 ºC (Figure 3b).Schulz scans yield the frequency density of crystal orientations with respect to the AAO nanopore axes and the PVDF nanotube axes - or equivalently the frequency density of the magnitude of an angle ψ between the PVDF nanotube axes as well as the AAO nanopore axes on the one hand and the reciprocal lattice vectors belonging to specific sets of lattice planes on the other hand. During a Schulz scan, the AAO membrane containing PVDF nanotubes aligned in its nanopores is rotated by the angle ψ about an axis, which is the intersection of the surface of the AAO template and the scattering plane (defined by incident wave vector and scattering vector). After crystallization at 136°C, the (110) and (020) lattice planes show almost identical orientation frequency densities with a maximum ψ = 0º. This result indicates the existence of two populations of crystals, one of which has the (110) and one of which has the (020) planes preferentially oriented normal to the PVDF nanotube axes. As $T_c$ is increased to 140°C, the relative population of crystals having their (020) planes oriented normal to the PVDF nanotube axes increases, whereas the relative population of crystals having their (110) planes oriented normal to the PVDF nanotube axes decreases. Further increase of $T_c$ to 152°C results in a clear preferred orientation of confined crystals. The maximum intensities as function of ψ for the (110) and (021) reflections occur at ψ ~ 53º and ψ ~ 45º, respectively. Since these ψ angles correspond to the angles enclosed by the (020) planes and the (110) as well as the (021) planes, this outcome confirms the preferential orientation of the (020) planes normal to the AAO nanopore axes.

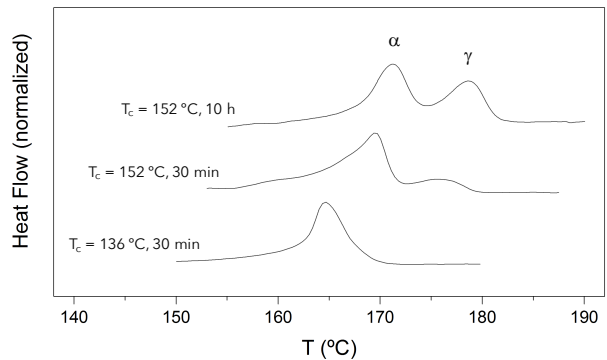

**Figure 4**. DSC heating scans for neat PVDF nanotubes crystallized at $T_c$ = 136 ºC for 30 min, $T_c$ = 152 ºC for 30 min and $T_c$ = 152 ºC for 10 h. Nanotubes crystallized at $T_c$ = 152 ºC exhibit two endothermic peaks corresponding to the melting of α- and γ-crystals, as indicated above the curve.

$T_c$ has also a striking impact on the crystallographic phase formed in the PVDF nanotubes. We found that after extended annealing periods a high $T_c$ polar γ-crystals develop, hence adding ferroelectric properties to the PVDF nanotubes. Figure 4 shows the DSC heating runs for PVDF nanotubes crystallized at $T_c$ = 136 ºC for 30 min, $T_c$ = 152 ºC for 30 min, and $T_c$ = 152 ºC for 10 h. Whilst the curve for nanotubes crystallized at 136 º displays a single endothermic peak that relates the melting of α-PVDF crystal, both curves for nanostructures crystallized at 152 ºC exhibit, apart from the melting peak of α-PVDF crystals, an additional high-temperature endothermic process that is associated with the melting of the more stable γ-PVDF crystals [23-25]. Moreover, this endothermic peak is more intense when PVDF nanotubes are crystallized for a longer period. This might suggest that during high-temperature annealing γ-crystals develop via transformation from α-crystals, as proposed by Lovinger[24]. We note, however, that bulk PVDF subject to the same thermal treatment retained the non-polar α-form (Supporting Information S7), suggesting that spatial constraint favors the polymorphic transformation. We note that the detection of γ-PVDF crystals in the WAXS patterns shown in Figure 3 is not possible, as the (020) and the (110) reflections of both forms appear at very similar 2θ values.

4.  **Conclusions.**

In conclusion, our results demonstrate that the crystallization process and the crystalline features (polymorph, texture, etc.) of nanoconfined polymers can be readily manipulated employing simple strategies adapted from methods used for bulk polymers. Supramolecular nucleating agents can be employed to modify the crystallization process of PVDF nanotubes. Moreover, isothermal crystallization allows controlling the texture and the crystals form of PVDF nanostructures. Interestingly, we show that high crystallization temperatures and long annealing periods induce the formation of uniaxially oriented γ-crystals, hence enabling the simple production of ferro/piezoelectric nanostructures. Clearly, the benefits of the approaches presented here are not limited to the materials above, thereby opening up a plethora of new possibilities for the processing and structure control of polymer-based nanostructures, including modulating their crystallization rate, tailoring crystal dimensions, tuning crystal orientation as well as selecting specific polymorphs that confer new properties and functionalities to the nanomaterial.

**Supporting Information:**

Details of Experimental Methods, DSC scans of bulk PVDF samples, details of isothermal crystallization analysis, calculation of the overall activation energy for crystallization, crystallinity of the PVDF materials, crystal orientation of PVDF nanotubes non-isothermally crystallized with and without BTA, DSC heating scans of bulk PVDF crystallized at low and high temperatures.

**Acknowledgments**

J.M. acknowledges support from the European Union's Horizon 2020 research and innovation programme under the Marie Skłodowska-Curie grant, agreement No 654682, the Provincial Council of Gipuzkoa under the programme Fellow Gipuzkoa and MICINN (FPU AP2005-1063). M.S. thanks the European Research Council (ERC-CoG-2014; project 646742 INCANA) for funding. N.S. is thankful for financial support from the research network Saxony-Anhalt 'Nanostructured Materials'. H.-W.S. is grateful to the German Research Foundation (DFG) for financial support within the Collaborative Research Center 840 (SFB 840, project B4). We acknowledge the support of Sandra Ganzleben in the synthesis of the BTA and Frank Abraham and Klaus Kreger for valuble discussions about the nucleating agents. Technical support from Dr. R. Benavente (DSC measurements) and D. Gómez (SEM), as well as from K. Sklarek and S. Kallaus (preparation of AAO) is also gratefully acknowledged.

**TOC graphic**

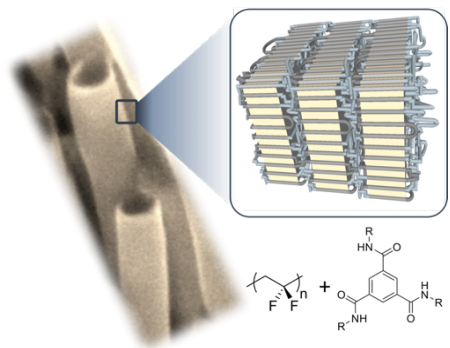